# Frequency-Based Patrolling with Heterogeneous Agents and Limited Communication

Tao Mao[1] and Laura E. Ray[1]
[1]Thayer School of Engineering, Dartmouth College, Hanover, NH 03755, United States

**Abstract -** *This paper investigates multi-agent frequency-based patrolling of intersecting, circle graphs under conditions where graph nodes have non-uniform visitation requirements and agents have limited ability to communicate. The task is modeled as a partially observable Markov decision process, and a reinforcement learning solution is developed. Each agent generates its own policy from Markov chains, and policies are exchanged only when agents occupy the same or adjacent nodes. This constraint on policy exchange models sparse communication conditions over large, unstructured environments. Empirical results provide perspectives on convergence properties, agent cooperation, and generalization of learned patrolling policies to new instances of the task. The emergent behavior indicates learned coordination strategies between heterogeneous agents for patrolling large, unstructured regions as well as the ability to generalize to dynamic variation in node visitation requirements.*

## 1 Introduction

A persistent patrolling task can be described as the activity of repeatedly visiting specified points in space in order to monitor the surrounding environment or perform a specified function at each point visited. The patrolled area can be described as a graph, in which each vertex or node represents a point to be visited and edges represent possible paths. The problem can be cast according to characteristics of real tasks that are modeled; for example, the number of agents, the number of nodes, and node connectivity may be specified according to patrolling needs and environment constraints. In prior work, it has generally been assumed that each node should be visited with the same frequency, and the resulting multi-agent planning problem is thus characterized as a combinatorial optimization problem with one of several possible criteria defined to maintain uniform visitation; however, it is not always the case that uniform visitation is desirable. Nodes can have different visitation requirements owing to their prominence, patrolled site structure, or activity level within the region [1].

The patrolling problem can be modeled as a coverage problem on closed or open polygons. In prior work, every point on the polygon is generally treated as having equal visitation requirements, and the objective is to either minimize the frequency variance or maximize the average/minimal frequency [2, 3]. The intuitive solution, based on finding a closed polygon for all agents to patrol in the same direction, guarantees uniform visitation and avoids conflicts between agents [2~4]. Elmaliach et al. further develop solutions to this uniform patrolling problem for open polygons [2, 3]. Their work allows agents to overlap trajectories in space but not in time and gives an analytical solution of this model considering real-world uncertainties.

Graph-based patrolling problems, which are abstracted from polygon continuous space patrolling, have also been proposed. In graph-based patrolling with uniform patrol rates for each node, the objective is to either maximize visitation frequency for every point [5] or minimize idle time for every point [6~8]. In models used in these papers, graph edges can have different lengths. For a single agent, the patrolling problem becomes a Travelling Salesman Problem [6]. In multi-agent cases, Hamilton circuit finding is given first priority [5, 6]; if Hamilton circuits do not exist, the algorithms in [5] find the longest path and include outliers, while partition-based strategies are applied in [6]. Reinforcement learning is applied in [7], which seeks an optimal graph partition for each member of the team to patrol. The graph model can be abstracted further in that patrolling efforts are focused on nodes, while edge lengths are ignored and synchronic node visits take place [9]. This abstraction is appropriate for problems involving patrolling of a large region, where the time spent at each node is substantial in comparison to travel time between nodes.

In this paper, we consider the problem of multi-agent patrolling in which the routes patrolled are modeled as undirected circle graphs that intersect occasionally, where one or more agents may patrol each route, and where communication between agents is sparse. As in [9], we focus on patrolled nodes, ignoring edge lengths and assuming synchronic node visitation. The problem is motivated by the need for coordination strategies between heterogeneous agents in patrolling large, unstructured regions. For example, in littoral patrolling, unmanned ground vehicles are constrained to operate on coastal roads, while unmanned robotic boats may operate on the surface of the water, and unmanned submersibles operate at depths but can also surface. We assume arbitrary visitation frequency requirements for graph nodes.

An objective of solving the patrolling problem described above is to investigate robustness of patrolling policies to variations in the patrolling task, specifically the node visitation frequency requirements and the number of nodes on each graph. Additionally, we wish to examine dynamics that emerge when patrolling policies specific to a given graph model are applied to a variant of that model. For this reason, the problem is cast into a reinforcement learning (RL) framework [10]. The RL algorithm for the patrolling task requires little prior knowledge, and the learned policies are adaptable to the patrolling environment, static or dynamic. However, like other combinatorial optimization methods, RL described as in [11] suffers from "the curse of dimensionality" as the number of nodes and/or agents increases. Thus, we use partially-observable states in order to manage the state space size. The reduction in state space size still permits convergence. Additionally, communication between agents is limited.

The paper is structured as follows: first, we formally define the multi-agent frequency-based patrolling problem targeted in this paper. Next, we review RL and partially-observable states, as applied to this problem. A policy generation method is developed; these policies are exchanged through sparse inter-agent communication to improve the team performance. Finally, we provide empirical results of convergence and communication rates for RL applied to multi-agent frequency-based patrolling. We also evaluate robustness and dynamics of patrolling scenarios in which learned policies generated under a static set of patrolling requirements are applied to a scenario with novel requirements due to variation in node visitation frequency requirements, or number of nodes.

## 2 Problem definition

**Definition 1** *Graph.* [12] A graph is an order pair $G = (V, E)$ consisting of a set of $n$ vertices (or nodes) $V=\{1, 2, 3…n\}$ and a set of $|E|$ edges: $E \subseteq V \times V$. An edge $e$ is a pair $e = (i, j) \in E$. A graph $G$ is undirected if $\forall\ e = (i, j) \in E \Leftrightarrow e' = (i, j) \in E$. A graph $G$ is directed if $\exists\ e = (i, j) \in E$ such that $e' = (i, j) \notin E$

In the patrolling problem, a vertex represents a patrolled node and an edge a possible path choice for the agent.

**Definition 2** *General multi-agent patrolling problem.* A group of $r$ agents patrols an area represented by an undirected graph $G = (V, E)$.

According to definition 2, there are $n$ nodes to be patrolled and $|E|$ possible paths for $r$ agents to move. We define the agent to be in position $X_i$ if it is currently visited node $i$. A *step* of a patrolling agent in patrolling is a move from one node to an adjacent one.

**Definition 3** *Circle.* [13] An undirected graph $G = (V, E)$ is considered a circle $C = (V_c, E_c)$ if it has only the set of edges $E= \{(i_1,i_2), (i_2,i_3)…(i_{n-1},i_n), (i_n,i_1)\}$. For simplicity, but without loss of generality, we denote the set of vertex in the circle as $V_c$, and the set of edges as $E_c = \{(1,2), (2,3)…(n-1,n), (n,1)\}$.

**Definition 4** *Multi-agent frequency-based patrolling.* A group of $r$ agents patrols an area represented by a graph that consists of one or more circles, where each node has a visitation frequency requirement, or patrol rate, $F_i$. One or more agents can patrol each circle.

The goal of the patrolling task is for the team of agents to minimize the differences between the actual visitation frequency $f_i$ and required frequency of visitation $F_i$ for each node. The patrol route may consist of a single circle graph or multiple, intersecting circle graphs. We abstract dynamics of the vehicles by assuming that the time spent at each node is large compared to the travel time between nodes by each agent. Agents are assumed to communicate only when they occupy either the same node or adjacent nodes so as to model realistic communication limitations in large, unstructured environments.

## 3 Reinforcement learning for multi-agent Frequency based patrolling

The RL problem, which is Markov Decision Process, can be described as a tuple $\langle R, S, A, p, \rho \rangle$, where $R$ is the set of learning agents; $S$ is the set of possible states; $A$ is the set of actions that the agents can choose from; $p : S \times A \times S \to [0,1]$ is the state transitional probability; $\rho : S \times A \times S \to \mathbf{R}$ is the reward function [10]. RL, Q-learning in particular, defines a state-action value $Q : S \times A \to \mathbf{R}$, which evaluates the value of a certain action taken under a state. Q-learning is a RL algorithm that aims to find satisfactory approximations for $Q$ values during the learning process and uses these $Q$ values to obtain the best action policies for a particular system. The state-action value $Q$ is updated at the end of each episode, and the learned value function $Q_t$ approximates the optimal value function as the algorithm iterates if conditions on the learning rate are met [14]. We use ε-greedy action-value method to determine next actions during learning. This method balances action policy exploration and exploitation.

In the multi-agent frequency-based patrolling problem described in the previous section, the agents can only estimate the frequencies of visitation of nodes and do not know the true values because they cannot obtain other agents' visitation histories except when communication is available. Also, if we choose a state space comprising *every* patrolled node, the dimension of the learning space increases exponentially with the number of nodes. Considering that each agent moves only to an adjacent node at each step, we propose a partially-observable state comprised of difference frequencies of four nodes nearest to the current node $i$ occupied by an agent $r$:

$$s_g \to s = \left[\Delta f^{i_{-2}}, \Delta f^{i_{-1}}, \Delta f^{i_1}, \Delta f^{i_2}\right] \quad (1)$$

$\Delta f^{i_{-2}}$, $\Delta f^{i_{-1}}$, $\Delta f^{i_1}$, and $\Delta f^{i_2}$ are the difference between the required visitation frequency and the actual visitation frequency for the four nodes nearest to the currently occupied node $i_0$. In eq. 1, $s_g$ is a global or environmental state while $s$ is a partially-observable state, which we use to model the Markov decision process and train the RL algorithm, for its state space is considerably smaller than that of a global state. At each time step, agents may move either to the left or right, i.e., there are two actions to choose from.

Each agent is provided with the target frequency requirements for all nodes at $t = 0$. During the learning phase, each agent maintains its own visitation history $H_j$. However, an agent does not know with certainty the visitation frequencies, histories, or policies of other agents. Communication assumptions described previously enable only sparse transmission of information between agents.

## 4 Policy generation and exchange via inter-agent communication

Since communication is sparse, the information transmitted when communication occurs is important. The state representing the environment is partially observable; hence, the agents should exchange their visitation histories as well as their patrolling policies in order to reduce uncertainty in estimating the aggregate visitation frequencies in eq. 1 at certain time steps.

The theoretical framework for RL is based on properties of Markov Decision Processes. Thus, it is natural to define a policy as a Markov process, which, in the patrolling scenario, is defined as a Markov chain, or a stochastic process with Markov properties. It is assumed that the process has a finite set of states, in which one state transits to another only depending on its current state. The state used in the context of a Markov chain is the position $X_i$ of an agent as in Def. 2, and not the concept of state used for RL. Given historical data, we model an agent's next state to be conditioned on the past states as first-order Markov chain:

$$\Pr\left(X^{t+1} = X_i \mid X\right) = \Pr\left(X^{t+1} = X_i \mid X^t\right) \quad (2)$$

$X$ is a series of states $\{X^t, X^{t-1},\ldots, X^1, X^0\}$. From this equation, we define the transition probability as

$$p_{ij} = \Pr\left(X^{t+1} = X_j \mid X^t = X_i\right) \quad (3)$$

A transition matrix $P$ for an agent is comprised of elements of probability $p_{ij}$. The agent updates its transition matrix $P$ from its visitation history using a window of length $w$ time steps. The initial calculation of $P$ is performed after the first $w$ time steps transpire.

A policy $\pi$ is a stationary policy if it satisfies $\pi = \pi P$ For computation purpose, a stationary policy $\pi$ can be calculated as

$$\pi = diag\left[\lim_{k \to \infty} P^k\right] \quad (4)$$

where the $diag[*]$ operator retrieves the diagonal elements of the matrix.

Since each agent maintains their own transition policy $\pi_r$, they can exchange policies and visitation histories when they communicate. The estimated visitation frequency of a given node can then be obtained both from the agent's own visitation history, other agent's histories when these are available through a prior communication, and the agent's beliefs of others' motions given policies $\pi_{-r}$ exchanged during communication. Note that when policies are exchanged, they are not necessarily steady-state policies.

At every step, an agent records its own current position, and estimates other agents' positions according to policies $\pi_{-r}$ exchanged from others. Using both true and estimated histories, the agent calculates visitation frequencies for all nodes. Since estimates of visitation frequencies are based on sampled data, a time window of $w_f$ steps is used to estimate these frequencies. Each agent estimates its current state $s_e$ based on other agents' histories $H_{-r}$ of visitation frequencies (as estimated from exchanged histories and policies $\pi_{-r}$) and its own recorded history $H_{-r}$:

$$s_e = \left[\Delta f_e^{i_{-2}}, \Delta f_e^{i_{-1}}, \Delta f_e^{i_1}, \Delta f_e^{i_2}\right] \quad (5)$$

$\Delta f_e^{i_{-2}}$, $\Delta f_e^{i_{-1}}$, $\Delta f_e^{i_1}$, and $\Delta f_e^{i_2}$ are the difference between the required visitation frequency and the *estimated* visitation frequency for the four nodes nearest to the currently occupied node $i_0$. Note that visitation histories $H_{-r}$ of other agents are initialized as null, and transitional matrices $P_{-r}$ are initialized as uniform prior to agent communication.

## 5 Results

This section presents empirical results based upon two patrolling configurations – a benchmark configuration comprised of a single circle route with uniform patrol frequency requirements (Fig. 1a) and a more general configuration comprised of three intersecting circle graphs, each having a different number of patrolled nodes and non-uniform patrol frequency requirements (Fig. 1b). Each node has a specified component frequency $C_i$, and

$$\sum_{i=1}^{n} C_i \leq 1 \quad (6)$$

Eq. 6 assures that the team has sufficient overall capacity to perform the patrolling task. Our notation for visitation frequency assumes that each agent has a capacity of 100%, and thus the total capacity is $100r\%$, where $r$ is the number of agents. Then, the visitation frequency requirement for Node $i$ is defined as $F_i = 100rC_i\%$.

Q-learning is used in the learning phase, with an ε-greedy criterion for action selection. ε is initialized as 1 and exponentially decays to 0.01. An agent receives a reward of $f_i - F_i$ when it occupies Node $i$.

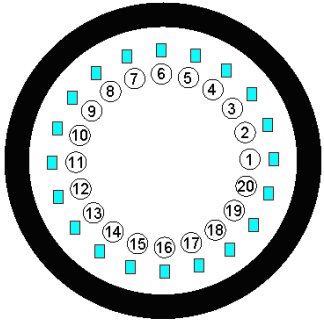

(a) Benchmark configuration

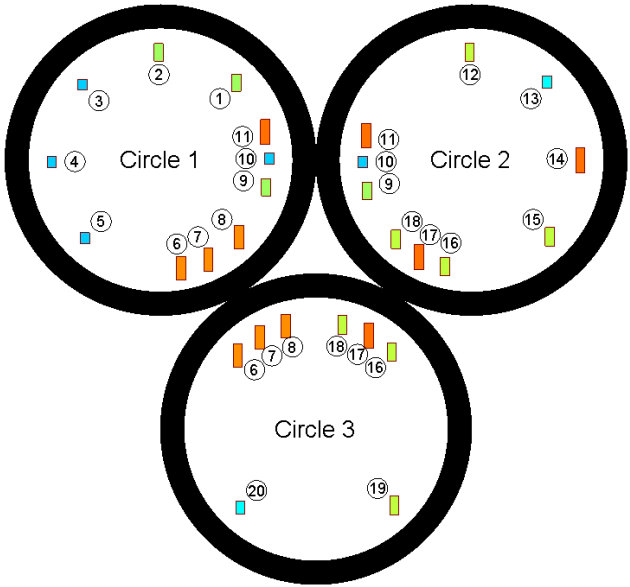

(b) General configuration

Figure 1. Patrolling configurations

Time window parameters $w$ and $w_f$ are 1000 and 100 steps respectively. We assess the efficiency of the algorithm by measuring the difference between the required and actual patrol rate at each node as a function of learning step. We formally define the insufficiency of the patrol rate $I_i$ for Node $i$ as a performance metric:

$$I_i = \begin{cases} F_i - f_i, & F_i - f_i > 0 \\ 0, & F_i - f_i \leq 0 \end{cases} \quad (7)$$

Note that this measure does not penalize the use of excess capacity, i.e., when the actual patrol frequency exceeds the desired patrol frequency, sufficiency requirements are met and thus insufficiency of patrol rate is zero.

## 5.1 Single-circle multi-agent patrolling

The benchmark configuration of Fig. 1a has $n = 20$ nodes, $r = 2$ patrolling agents, component frequency $C_i = 4.5\%$ for each node, and uniform patrolling frequency requirements of $F_i = 9\%$ at each node; thus, the system is overmanned in the sense that collectively, agents have sufficient capacity to meet visitation frequency requirements. This configuration is designed to test the learning algorithm in the case where there exist multiple, obvious solutions. Here, the range of policies that meet a uniform patrol frequency requirement varies: agents each traversing the route in a single direction, splitting the graph equally between agents, or agents each making random moves with uniform probability would all comprise solutions meeting this requirement. In the absence of additional criteria, all solutions are satisfying, or acceptable solutions. A RL approach generally converges to more stochastic than deterministic solutions when two actions are selected with equal probability during learning exploration.

We perform 10 experimental runs of $10^6$ learning steps each. In each run, agents learn a patrolling policy from scratch. Each learning step corresponds to a time step owing to abstraction of travel between nodes. Figure 2 shows mean and 1-sigma variation of $\sum I_i / n$ over 10 experimental runs and 20 nodes. Figure 2 also shows the number of instances of communication between two or more agents out of the last 1000 steps averaged over 10 experimental runs. These measures are shown as a function of learning step. Results show that a patrolling policy that achieves a visitation frequency within 1% of the target frequencies over 10 experimental runs emerges, providing a satisfying policy in about 200,000 steps. Three stages of convergence are evident. In the "adjustment" stage (< 10,000 steps), agents explore a new environment, collect experience for learning, and communicate sparsely. In the "convergence" stage (10,000 ~200,000 steps), learning starts to improve performance, reducing the average insufficiency of patrol rate. Finally, in the "converged" stage (> 200,000 steps), a satisfying policy emerges, and there is little room for the two-agent group to improve performance. Communication plays an important role in learning convergence as the increase in communication rate starting at ~10,000 steps helps the agents estimate their belief states with less uncertainty and makes Q-learning converge. In addition, the communication frequency has changed from random to more regular, as indicated by the communication pattern shown as the two small figures in Fig. 2. In the converged policy, each agent chooses to move left and right with nearly 50% frequency. Thus, each agent's policy comprises a roughly 50-50 chance of going left or right, which is a reasonable solution for this benchmark problem.

## 5.2 Multi-circle multi-agent patrolling

In the general configuration, 20 nodes are distributed in three circles for three agents to patrol, as in Fig. 1b. Some nodes are common to two circles and appear twice in the figure. Component frequencies for the 20 nodes are assigned non-uniformly, while obeying the constraint imposed by eq. 6. The component frequencies are 4.33%, 4.33%, 2.67%, 2.67%, 2.67%, 6.00%, 6.00%, 6.00%, 4.33%, 2.67%, 6.33%, 4.67%, 3.00%, 6.33%, 4.67%, 4.67%, 6.33%, 4.67%, 4.67%, 3.00%, and required visitation frequencies are three times component frequencies. The sum of component frequencies is 90%, indicating that the team has sufficient capacity. In Fig. 1b, the length of the bars inside the circles associated with each node indicates the resulting $F_i$, as does the color, with blue indicating lowest $F_i$ and red indicating highest $F_i$.

We perform 10 experimental runs of $10^6$ steps each, in which agents learn from scratch for each run. Figure 3 shows mean and 1-sigma variation in the mean insufficiency of patrol rate averaged over 20 nodes and 10 experimental runs, as well as the number of instances of communication between two or more agents out of the last 1000 steps averaged over 10 experimental runs, as a function of learning step. A satisficing policy that achieves average insufficiency of patrol rate below 1% is learned in ~$10^5$ steps. As in the benchmark configuration, there are three stages in convergence of the learning algorithm. Improved convergence rate over the benchmark problem is attributed to the lower number of patrolled nodes per agent.

Figure 4 shows snapshots of the visitation policy convergence for each patrolling agent in the form of a "heat map" designating the frequency of occupancy of each agent on the graph and a colored bar for each node. The color in a bar indicates the visitation requirement as specified in Fig. 1b. The length of color filled in each bar denotes the patrol rate agents provide to a node in order to fullfil the frequency requirement. For overlapped nodes, the sum of each agent's contribution to patrolling a node, denoted by the color of the associated point in the circle, produces the total actual visitation frequency. Prior to convergence, visitation requirements are not satisfied at every node, and agents do not exhibit strong cooperation. Subsequently, the agents begin to increase their communication rate as seen in Fig. 3; cooperation emerges; and the patrolling requirements are met. For instance, Agent 1 has the heaviest workload along its circle (11 nodes) and Agent 3 has a comparably lighter workload (8 nodes). Thus, in the overlapped nodes on these two graphs, Agent 3 takes most of the workload. Policy convergence is achieved with cooperation.

## 5.3 Dynamic multi-agent patrolling

Due to the number of learning steps required to establish a patrolling policy, it is desirable that a learned policy generalize to variations in node visitation requirements and number of nodes. Here, we evaluate the transient dynamics and steady-state performance when the learned policy is applied to each of two variants of the general configuration. In the first variant, frequency requirements for Node 4 and Node 14 are swapped, with Node 4 requirements increasing from from 8% to 19% and Node 14 decreasing from 19% to 8%; and in the second, a scenario of adding and deleting a patrol node is obtained by setting the frequency requirement for Node 2 from 13% to zero and adding a new node Node 21 between Node 19 and 20 with a frequency requirement 14%.

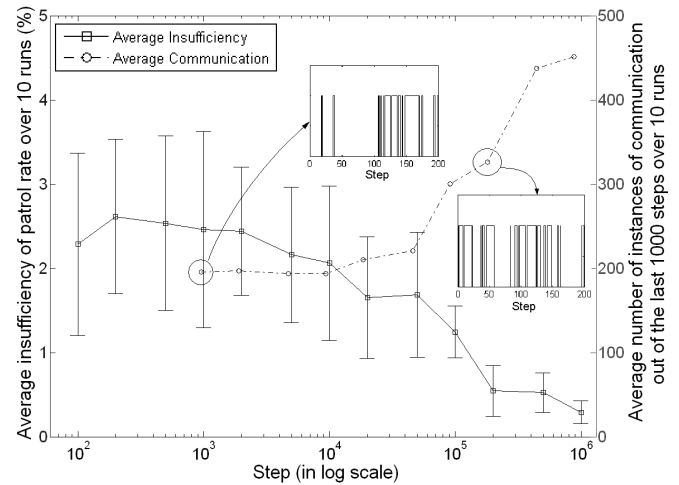

Figure 2. Learning convergence of benchmark configuration

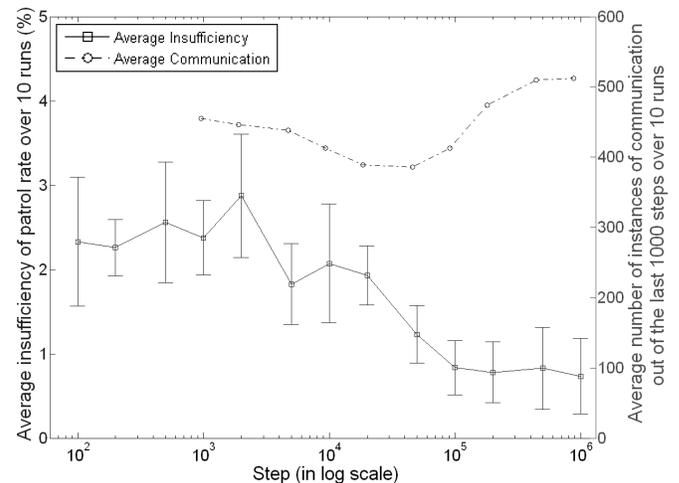

Figure 3. Learning convergence of general configuration

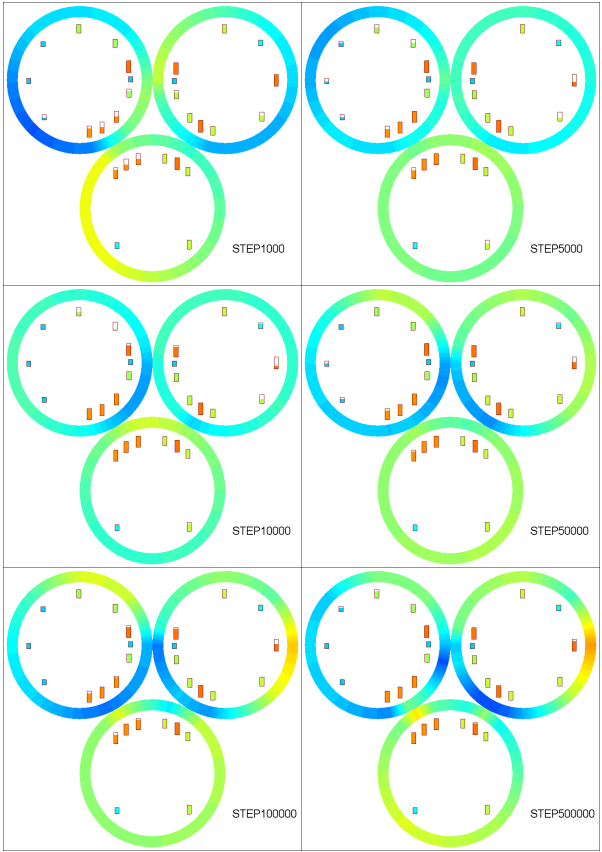

Figure 4. Snapshots of learning in the general configuration

Figure 5 indicates transients in $D_i = f_i - F_i$, the difference between actual and required visitation frequency at each node for each experiment. In the first experiment (Fig. 5a), the agents reduce the frequency difference range from between -11.1% and 11.4% to between -2.1% and 5.3% within 100 steps. After 100 steps, most transients decay to within in the color range of green, yellow or red, indicating frequency requirements of these nodes are met or exceeded. In the second experiment (Fig. 5b), the agents reduce the frequency difference range from between -14% and 13.4% to between -3.1% and 4.7% in 100 steps. After 100 steps, most transients decay to within in the color range of green, yellow or red, which indicates frequency requirements of these nodes are met. These examples indicate that the collective learned policies can be applied to a new instance of the task in a dynamic environment with new nodes.

Figures 6 and 7 show dynamic evolution of the heat map for each of these experiments. In Fig. 6, cooperation emerges between Agents 1 and 2; due to workload reduction for Node 14, Agent 2 increases visitation of Node 11 to help Agent 1 meet the increased demand from Node 4. In Fig. 7, cooperation emerges between three agents: Agent 1 assists Agent 3 by assisting Agent 2. Due to zero workload for Node 2, Agent 1 applies effort to visit Nodes 9~11; then, Agent 2 has the capacity to help Agent 3 alleviate the workload on Nodes 16~18; finally, Agent 3 can handle new Node 21. In other words, indirect cooperation between Agent 1 and Agent 3 emerges through Agent 2's involvement. Node 2, which requires no visits, is not avoided due to its physical existence in Circle 1; however, the visitation frequency for this node is reduced to 4%.

Comparison of steady-state heat maps for dynamic variations of the task from Fig. 6 and 7 with the heat map for the standard configuration at convergence of learning (Fig. 4, snapshot at step 500,000) shows that a range of dynamics can be elicited from a single-run learned policy.

# 6 Conclusions

We apply Q-learning based on partially-observable states to a multi-agent frequency-based patrolling problem where frequency requirements for each node can be non-uniform, and communication is limited. Transition matrices describing the probability of state transitions are derived from Markov chains, and policies are exchanged when two agents occupy the same or adjacent nodes. In both a benchmark configuration with uniform patrol frequency requirements and a multi-graph patrol configuration with non-uniform requirements, multi-agent teams learn patrolling policies embodying cooperation. Empirical results demonstrate that the patrolling policies learned from static configurations can be applied to new instances of the patrolling task, with variations in the number of nodes or changes in patrol frequency requirements.

The relaxation of uniform frequency requirements and modeling of communication as sparse makes the model realistic for patrolling large, unstructured regions using heterogeneous agents. Additionally, the more stochastic policies derived from RL can make it difficult for external agents to predict patrolling agents' behaviors, which is a desirable feature for many applications.

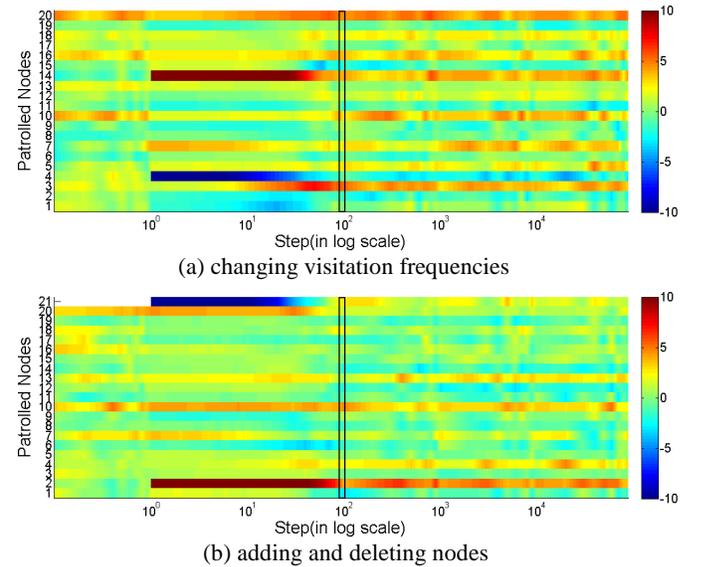

Figure 5. Transients in difference between actual and required visitation frequency at each node in dynamic environments

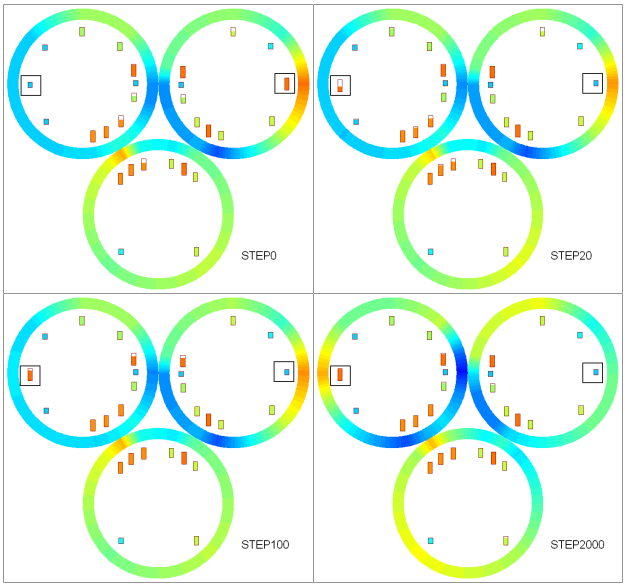

Figure 6. Policy transition in dynamic environment with changing visitation frequencies

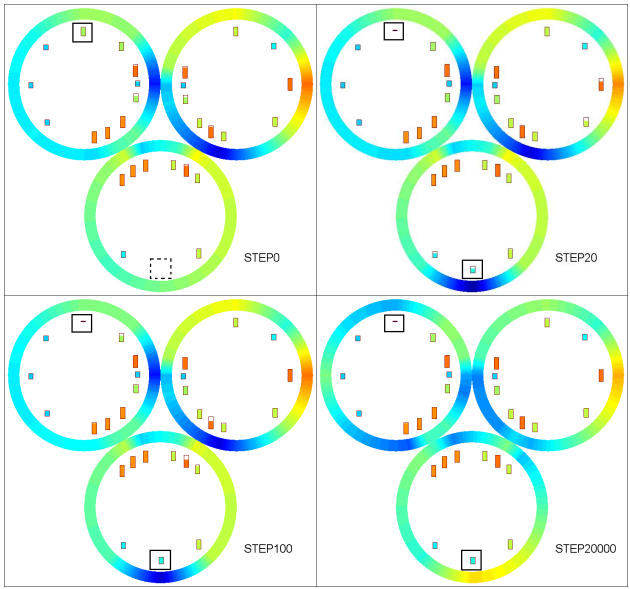

Figure 7. Policy transition in dynamic environment with adding and deleting nodes

# 7 Acknowledgements

This work is supported by the Office of Naval Research under Multi-University Research Initiative (MURI) Grant No. N00014-08-1-0693.